\documentclass[reprint,prl,aps, floatfix]{revtex4-1}
\usepackage{graphicx}
\usepackage{color}
\usepackage{dcolumn}
\usepackage{bm}
\usepackage{amssymb}
\usepackage{braket}
\usepackage{color,amsmath}
\usepackage{comment}
\usepackage{gensymb}
\usepackage{soul,xcolor} %use \st{text} to achieve strikethrough
\setstcolor{red} %set colour of strikethrough to red
\usepackage{epstopdf}
\usepackage{hhline}
\usepackage{tabularx}
\usepackage{xspace}
\usepackage{float}

\newcolumntype{C}[1]{>{\centering\arraybackslash}m{#1}}
\newcolumntype{N}{@{}m{0pt}@{}}

\definecolor{cadmiumgreen}{rgb}{0.0, 0.42, 0.24}

\newcommand{\moire}{moir\'e\xspace}

\begin{document}
% \linenumbers

\title{Mixed-dimensional \moire systems of graphitic thin films with a twisted interface}

\author{Dacen Waters$^{1,2*}$}
\author{Ellis Thompson$^{1*}$}
\author{Esmeralda Arreguin-Martinez$^{3}$}
\author{Manato Fujimoto$^{4}$}
\author{Yafei Ren$^{3}$}
\author{Kenji Watanabe$^{5}$} 
\author{Takashi Taniguchi$^{6}$} 
\author{Ting Cao$^{3}$}
\author{Di Xiao$^{3,1}$}
\author{Matthew Yankowitz$^{1,3\dagger}$}

\affiliation{$^{1}$Department of Physics, University of Washington, Seattle, Washington, 98195, USA}
\affiliation{$^{2}$ Intelligence Community Postdoctoral Research Fellowship Program, University of
Washington, Seattle, Washington, 98195, USA}
\affiliation{$^{3}$Department of Materials Science and Engineering, University of Washington, Seattle, Washington, 98195, USA}
\affiliation{$^{4}$Department of Physics, Osaka University, Osaka 560-0043, Japan}
\affiliation{$^{5}$Research Center for Functional Materials,
National Institute for Materials Science, 1-1 Namiki, Tsukuba 305-0044, Japan}
\affiliation{$^{6}$International Center for Materials Nanoarchitectonics,
National Institute for Materials Science, 1-1 Namiki, Tsukuba 305-0044, Japan}
\affiliation{$^{*}$These authors contributed equally to this work.}
\affiliation{$^{\dagger}$ myank@uw.edu (M.Y.)}

\maketitle

\textbf{Moir\'e patterns formed by stacking atomically-thin van der Waals crystals with a relative twist angle can give rise to dramatic new physical properties~\cite{Balents2020,Andrei2020}. The study of moir\'e materials has so far been limited to structures comprising no more than a few vdW sheets, since a moir\'e pattern localized to a single two-dimensional interface is generally assumed to be incapable of appreciably modifying the properties of a bulk three-dimensional crystal. Layered semimetals such as graphite offer a unique platform to challenge this paradigm, owing to distinctive properties arising from their nearly-compensated electron and hole bulk doping. Here, we perform transport measurements of dual-gated devices constructed by slightly rotating a monolayer graphene sheet atop a thin bulk graphite crystal. We find that the moir\'e potential transforms the electronic properties of the entire bulk graphitic thin film. At zero and small magnetic fields, transport is mediated by a combination of gate-tunable moir\'e and graphite surface states, as well as coexisting semimetallic bulk states that do not respond to gating. At high field, the moir\'e potential hybridizes with the graphitic bulk states owing to the unique properties of the two lowest Landau bands of graphite. These Landau bands facilitate the formation of a single quasi--two-dimensional hybrid structure in which the moir\'e and bulk graphite states are inextricably mixed. Our results establish twisted graphene-graphite as the first in a new class of mixed-dimensional moir\'e materials.}

Twisting two sheets of monolayer graphene by a small angle results in the formation of a long-wavelength moir\'e potential that substantially alters the low-energy bands~\cite{Bistritzer2011,Morell2010}. The moir\'e bands become extremely flat and isolated near the ``magic angle'' of approximately $1.1^{\circ}$, generating an array of strongly correlated states including magnetism and superconductivity~\cite{Cao2018a,Cao2018b,Lu2019,Yankowitz2019,Balents2020,Andrei2020}. Moir\'e flat bands also form upon incorporating additional graphene sheets into the structure, recently observed in the magic-angle trilayer/tetralayer/pentalayer family~\cite{Park2021,Hao2021,Park2022,Burg2022,Zhang2022} as well as in twisted monolayer-bilayer~\cite{Chen2021,Polshyn2020,Shi2021,He2021tmbg} and bilayer-bilayer graphene~\cite{Shen2020,Liu2020,Cao2020,Burg2019,He2021tdbg}. So far, the study of twisted graphene structures has mostly been limited to those assembled from monolayer and bilayer graphene building blocks, since thicker Bernal-stacked constituents contribute additional bands at low energy. Band structure modeling indicates that moir\'e bands are likely to persist to arbitrarily thick Bernal graphite structures, but remain localized at the twisted interface and coexist with conventional bulk graphite bands~\cite{Cea2019}. Moir\'e surface states have been observed previously in scanning tunneling microscopy experiments performed on highly oriented pyrolitic graphite with a rotationally faulted surface sheet~\cite{Li2010}. However, it is currently not know whether and how these moir\'e surface states impact the electronic properties of the entire bulk graphitic thin film.

%%%%%%%%%%%%%%%%%%%%%%%%%%%%%%%%%%%%%%%%%%%%%%%%%%%%%%%%%%%%%%%%%%%%%
\begin{figure*}[t]
\includegraphics[width=\textwidth]{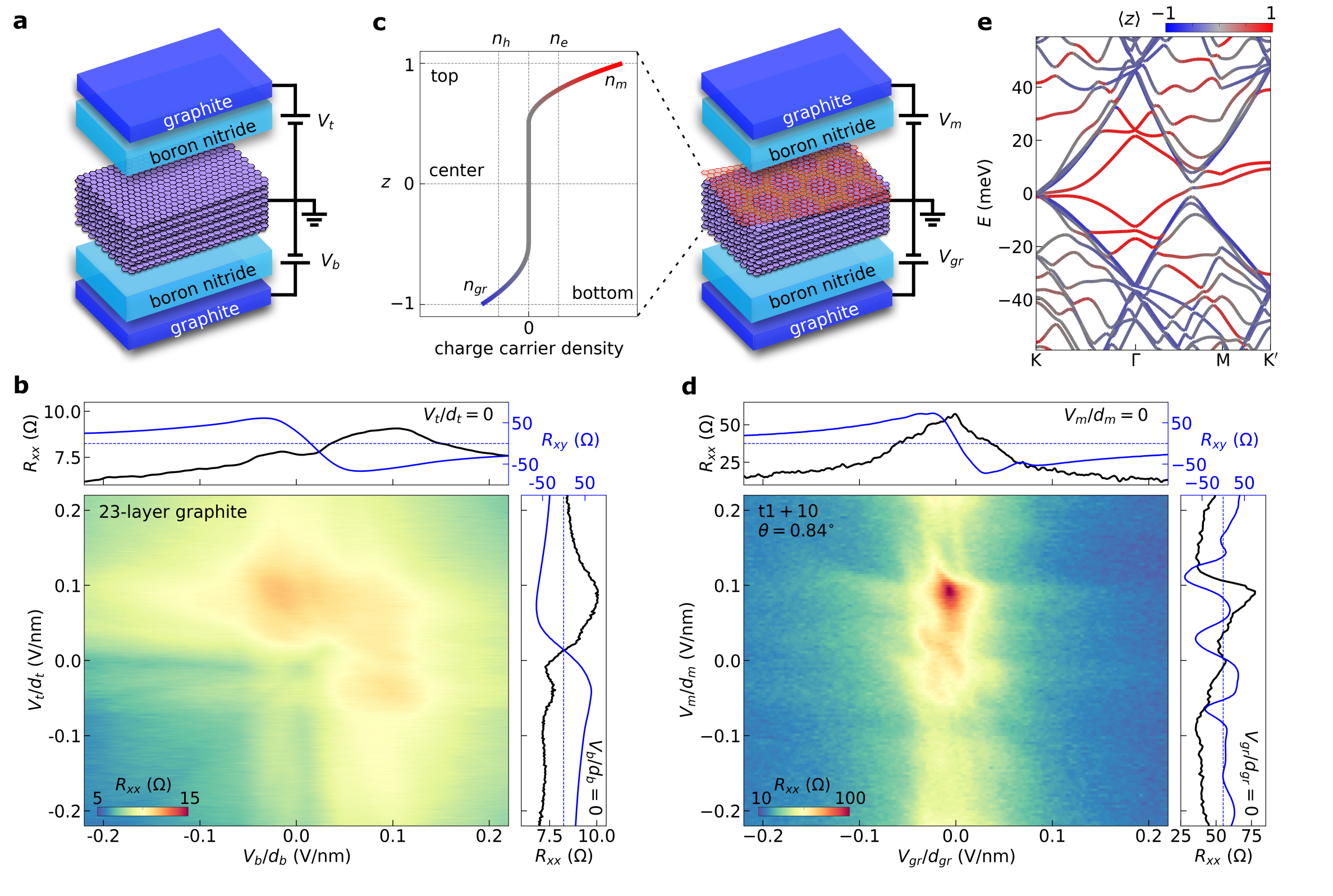} 
\caption{\textbf{Comparison of Bernal and moir\'e graphite at zero field.}
\textbf{a}, Cartoon schematic of a Bernal graphite thin film device with top ($V_t$) and bottom ($V_b$) gates.
\textbf{b}, Resistance of a 23-layer Bernal graphite thin film as a function of the top and bottom gate voltages. The top (bottom) gate voltage is normalized by the top (bottom) BN thickness, $d_t$ ($d_b$). The panel above (to the right) of the color map shows cuts of $R_{xx}$ and $R_{xy}$ as $V_b$ ($V_t$) is swept with $V_t=0$ ($V_b=0$). $R_{xy}$ is acquired with $B=0.2$~T. 
\textbf{c}, (Right) Cartoon schematic of a moir\'e graphite thin film device. The gate facing the moir\'e (Bernal graphite) surface is denoted $V_m$ ($V_{gr}$).
(Left) Schematic illustration of the free charge distribution in a dual-gated graphitic thin film as a function of the position along the $c$-axis of graphite (denoted as $z$). $n_m$ and $n_{gr}$ are the charge carrier densities on the \moire and Bernal graphite surfaces, respectively. These can be modified by tuning $V_m$ and $V_{gr}$, shown here with arbitrarily chosen magnitudes and signs. $n_e$ and $n_h$ denote the density of electron and hole carriers in graphite, which are approximately equal such that the total doping is zero in the bulk. The doping in the bulk does not depend on gating.
\textbf{d}, Resistance of a t1+10 graphite sample with $\theta=0.84^{\circ}$ as a function of $V_m$ and $V_{gr}$, normalized by the appropriate BN thicknesses. The panels above and to the right of the color map show cuts of $R_{xx}$ and $R_{xy}$, analogous to those in \textbf{b}.
\textbf{e}, Calculation of the band structure of a t1+10 graphite sample with $\theta=0.84^{\circ}$. The cut is taken along a contour within the moir\'e Brillouin zone. The color of bands corresponds to their expectation value along the graphite $c$-axis, denoted as $\langle\rm{z}\rangle$, where a value of 1 (-1) corresponds to the moir\'e (Bernal graphite) surface as shown in \textbf{c}.
}
\label{fig:1}
\end{figure*}
%%%%%%%%%%%%%%%%%%%%%%%%%%%%%%%%%%%%%%%%%%%%%%%%%%%%%%%%%%%%%%%%%%%%%

Here, we investigate the transport properties of graphitic structures with a moir\'e interface created by a single rotational fault within the crystal. We primarily focus on the case where the moir\'e potential is localized to one surface of the structure, achieved by rotating a flake of monolayer graphene by a small twist angle atop a Bernal graphite thin film. We also compare to the case where the moir\'e interface is buried at the center of the graphitic structure. We show that a single two-dimensional moir\'e interface can strongly modify the properties of the entire graphitic thin film, owing to a number of unique properties arising from its semimetallic nature. At zero and small magnetic fields, we find that the electronic transport can be well approximated by parallel contributions of gate-induced surface accumulation layers and intrinsic bulk states. In the case where the twisted interface is located at an outer surface of the sample, the transport properties are dominated by moir\'e surface bands and differ considerably from Bernal graphite. At high magnetic fields, the moir\'e and bulk states become inextricably mixed by a standing wave formed along the $c$-axis of graphite. The entire graphitic thin film becomes quasi--two-dimensional in this regime, forming a novel mixed-dimensional moir\'e system.

%%%%%%%%%%%%%%%%%%%%%%%%%%%%%%%%%%%%%%%%%%%%%%%%%%%%%%%%%%%%%%%%%%%%%
\begin{figure*}[t]
\includegraphics[width=6.9 in]{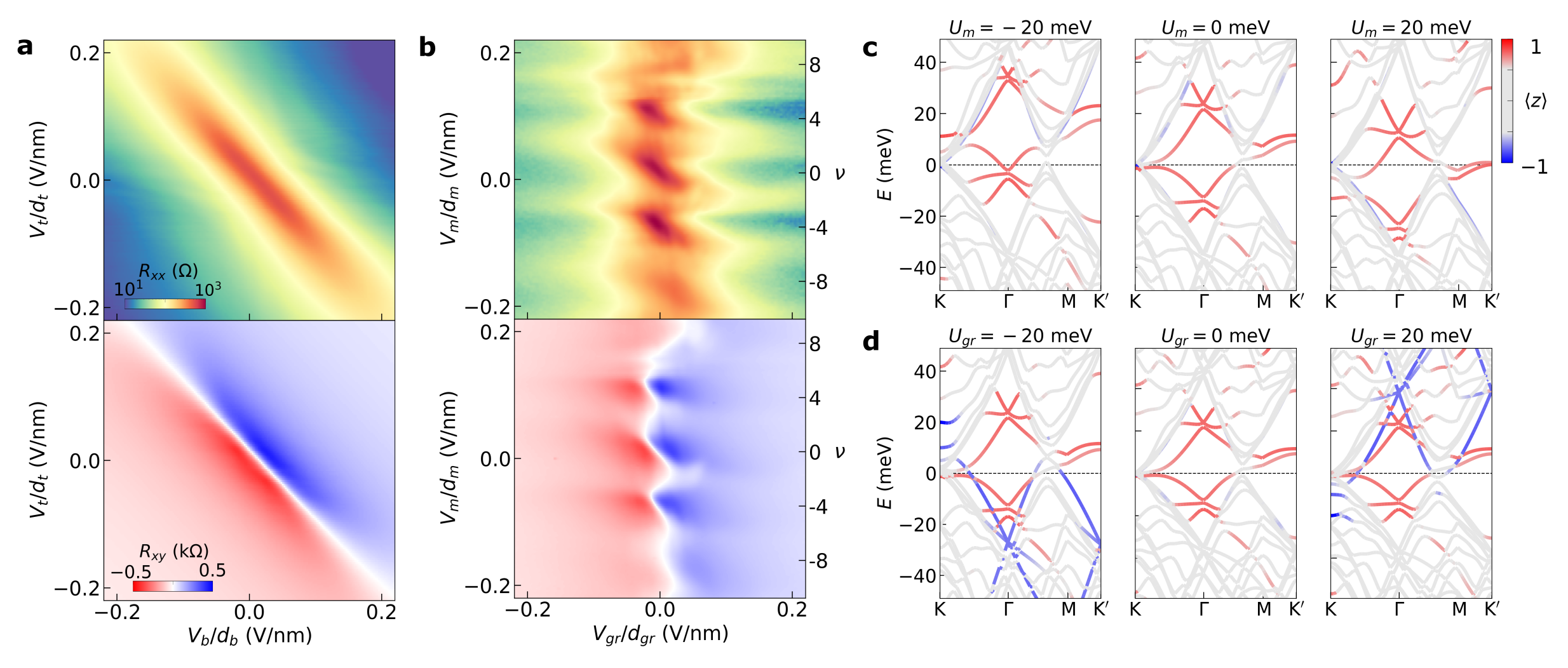} 
\caption{\textbf{Low-field magnetotransport and independent gating of surface-localized states.}
\textbf{a}, Longitudinal (top) and Hall (bottom) resistance maps of the 23-layer Bernal graphite device acquired at $B=0.5$~T. 
\textbf{b}, The same maps acquired in the t1+10 moir\'e graphite device. The vertical axis on the right indicates the filling factor, $\nu$, of the moir\'e surface bands.
\textbf{c}, Calculation of the band structure of a t1+10 graphite sample with $\theta=0.84^{\circ}$ as the surface potential at the moir\'e interface is changed from $U_m=-20$~meV (left) to $0$ (center) to $+20$~meV (right). The potential at the Bernal graphite surface is $U_{gr}=0$. The Fermi energy is held fixed at zero energy (black dashed line). The potential drop across graphite layers is modeled using a Thomas-Fermi screening approximation (see Methods).
\textbf{d}, The same calculation but with $U_m=0$ and a varying $U_{gr}$. In \textbf{c-d}, the bands are color-coded according to the expectation value of their $z$ coordinate along the $c$-axis of the graphitic structure.  
}
\label{fig:2}
\end{figure*}
%%%%%%%%%%%%%%%%%%%%%%%%%%%%%%%%%%%%%%%%%%%%%%%%%%%%%%%%%%%%%%%%%%%%%

\medskip\noindent\textbf{Comparison of transport in Bernal and moir\'e graphite}

We focus our study primarily on twisted graphene-graphite (i.e., t1+$Z$, where $1$ indicates monolayer graphene and $Z$ corresponds to the number of graphene sheets in the bulk thin film). We measure devices with graphite thickness varying from $Z=6$ to $40$ layers, and with twist angles between $\theta=0.84^{\circ}$ and $1.31^{\circ}$. These structures are encapsulated between flakes of boron nitride (BN), and capped by additional graphite flakes acting as top and bottom gates (see Methods for additional details). All transport measurements are performed at a base temperature of 1.7~K, unless otherwise noted. We first compare the transport properties of Bernal graphite with a representative twisted graphene-graphite device. Figure~\ref{fig:1} shows this for the case of a 23-layer Bernal graphite device and a t1+10 device with $\theta=0.84^{\circ}$. A schematic of the Bernal graphite device is shown in Fig.~\ref{fig:1}a, with top and bottom gate voltages denoted as $V_t$ and $V_b$, respectively. 

The color map in Fig.~\ref{fig:1}b shows the longitudinal resistance, $R_{xx}$, of the device at zero magnetic field as a function of the voltage on both gates (each gate voltage is normalized by the corresponding BN dielectric thickness). The resistance changes by only a few ohms with gating, consistent with the expectation that the gates are only able to dope the outer few layers of the nearest graphite surface due to screening in the bulk. Consequentially, the primary resistance features we observe evolve either vertically or horizontally in the map. Although the resistance map exhibits fine structure that we do not fully understand, we find that transport is very similar upon sweeping each gate with the other held fixed. This behavior is anticipated from the mirror symmetry of Bernal graphite, with small differences likely resulting from a variation in mobility between the two surfaces. This can also be seen by comparing the black traces in the panels above and to the right of the resistance map, which show $R_{xx}$ as each gate is swept with the other held at ground. We further see corroborating behavior in the Hall resistance, $R_{xy}$, measured in a small magnetic field of $B=0.2$~T. In particular, $R_{xy}$ exhibits a sign change around zero bias in each gate sweep, signifying a corresponding sign change in the charge of the free carriers residing in the surface accumulation layer.

In contrast to our observations in Bernal graphite, transport in our t1+10 sample differs considerably depending on which gate is swept, as the mirror symmetry of the structure is broken by the rotated graphene sheet at the surface. Here, we denote the voltage on the gate facing the moir\'e (Bernal graphite) surface as $V_m$ ($V_{gr}$) (see Fig.~\ref{fig:1}c). We see a much larger change in the resistance with gating in this device (Fig.~\ref{fig:1}d), with the highest resistance confined to a small region around $V_{gr}\approx0$. Transport is reminiscent of Bernal graphite when sweeping $V_{gr}$ (see $R_{xx}$ and $R_{xy}$ traces in the top panel), but exhibits fundamentally new behavior when sweeping $V_m$ (right panel). In particular, repeated $R_{xy}$ sign changes indicate multiple instances in which the free carriers on the twisted surface switch between electron- and hole-like. This behavior arises from the moir\'e reconstruction of the graphite band structure, marking the formation of a series of surface-localized moir\'e bands. These can be seen in calculations of the band structure of this material (Fig.~\ref{fig:1}e, see Methods for details). The bands are color-coded based upon their expectation value along the graphite $c$-axis, denoted as $\langle \rm{z} \rangle$, where a value of 1 (-1) corresponds to the moir\'e (Bernal graphite) surface. We find moir\'e bands localized on the outer graphene layers at the rotated interface (red colored bands), consistent with previous calculations performed for infinitely thick graphite slabs~\cite{Cea2019}. 

Despite the differences between the Bernal and moir\'e graphite devices, the gate-dependent transport of both are dominated by their surfaces. Figure~\ref{fig:1}c illustrates this schematically for the twisted graphene-graphite device, where a charge carrier density of $n_m$ ($n_{gr}$) can be induced on the moir\'e (Bernal) surface by tuning $V_m$ ($V_{gr}$). The graphite bulk remains a compensated semimetal, with roughly equal electron- and hole-doping ($n_e$ and $n_h$) that does not change with gating.

\medskip\noindent\textbf{Low-field magnetotransport properties of moir\'e graphite}

The contrast between Bernal and moir\'e graphite becomes more obvious upon applying a small magnetic field along the $c$-axis. Figure~\ref{fig:2}a shows a dual-gate resistance map for the 23-layer Bernal graphite sample acquired at a magnetic field of $B=0.5$~T. The resistance is largest when the voltage on both gates is approximately zero, consistent with prior reports of a large magnetoresistance (MR) in bulk graphite crystals~\cite{Soule1958}. Upon gating, we additionally find a large MR everywhere along the condition of overall charge neutrality, $V_t/d_t+V_b/d_b=0$, as evidenced by the diagonal resistance stripe in Fig.~\ref{fig:2}a. Measurements of the Hall resistance show a corresponding sign change across the line of overall neutrality. 

%%%%%%%%%%%%%%%%%%%%%%%%%%%%%%%%%%%%%%%%%%%%%%%%%%%%%%%%%%%%%%%%%%%%%
\begin{figure*}[t]
\includegraphics[width=6.9 in]{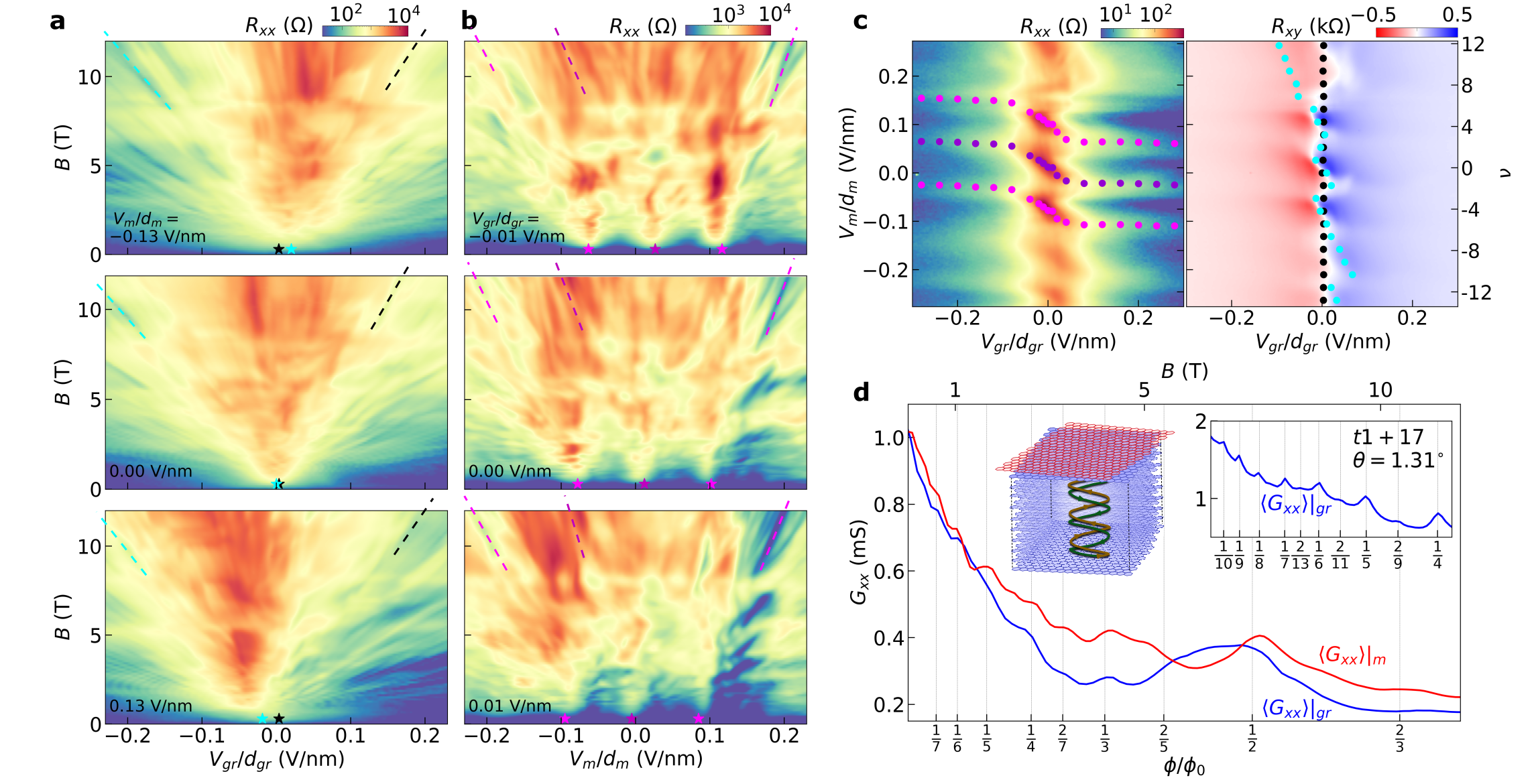} 
\caption{\textbf{Hybridization of moir\'e and bulk graphite states at high field.}
\textbf{a}, Landau fan diagrams from the t1+10 device acquired by sweeping $V_{gr}$ at the denoted values of $V_{m}$. The black dashed lines indicate selected QOs that project to $V_{gr}\approx0$ at $B=0$, whereas the blue dashed lines denote QOs that project to a value of $V_{gr}$ that changes with $V_m$. 
\textbf{b}, Landau fan diagrams acquired by sweeping $V_m$ at the denoted values of $V_{gr}$. The purple dashed lines indicate selected QOs that project to $\nu=0$, whereas the pink dashed lines indicate QOs that project to $\nu=\pm4$. 
\textbf{c}, Longitudinal (left) and Hall (right) resistance maps acquired at $B=0.5$~T. Zero-field projections of the $\nu=0$ and $\pm4$ states from the \moire gate Landau fans are overlayed on the longitudinal resistance map. Zero-field projections of QOs from the graphite gate Landau fans are overlayed on the Hall resistance map. Supplementary Videos 1-2 show the Landau fan diagrams used to extract all of the data points.
\textbf{d}, Conductance, $G_{xx}$, as a function of magnetic field. The blue (red) line is averaged over all values of $V_{gr}$ ($V_m$) for the Landau fan corresponding to the top panel in \textbf{a} (\textbf{b}). Selected rational values of the magnetic flux filling of the \moire unit cell, $\phi/\phi_0$ are denoted by vertical gray lines. 
The top right inset shows $G_{xx}$ averaged over all values of $V_{gr}$ for a $t1+17$ device with $\theta=1.31^{\circ}$ (see Extended Data Fig.~\ref{fig:t1_17} for the complete data set from this device).
The top left inset shows a cartoon schematic illustrating the formation of a standing wave in the lowest Landau bands of graphite (formed by the stacked blue graphene sheets) at high field. The standing wave hybridizes the Bernal graphite bulk states with the moir\'e surface states. The moir\'e interface is indicated by the rotated graphene sheet colored in red.
}
\label{fig:3}
\end{figure*}
%%%%%%%%%%%%%%%%%%%%%%%%%%%%%%%%%%%%%%%%%%%%%%%%%%%%%%%%%%%%%%%%%%%%%

These observations can be captured by a four-carrier Drude transport model, in which we assume that transport is mediated by coexisting electron and hole bulk carriers that do not respond to gating, along with gate-modulated surface accumulation layers controlled independently by the top and bottom gates (see Methods for full details). Extended Data Figure~\ref{fig:graphite_theory}a shows the result of this calculation, taking the known mobility and bulk charge doping of graphite~\cite{Soule1958,McClure1958}. The excellent agreement between our measurements and this calculation indicates that transport is primarily determined by the total free charge density in the material, even when the gate-induced charge is mostly localized at the outer surfaces. This behavior arises as a consequence of the low intrinsic bulk doping of graphite, as similar calculations performed with orders-of-magnitude larger bulk doping show virtually no gate tunability (Extended Data Fig.~\ref{fig:graphite_theory}b).

Corresponding measurements of the moir\'e sample reveal a more complex dependence of transport on gating (Fig.~\ref{fig:2}b and Extended Data Fig.~\ref{fig:t1_10_lowB}). Rather than a single resistance stripe, $R_{xx}$ exhibits a maximum that evolves with a peculiar zig-zag trajectory upon gating. The contour of $R_{xy}=0$ tracks closely with the $R_{xx}$ maximum. Notably, the periodic resets we observe upon biasing $V_{m}$ correspond closely with integer multiples of the gate voltage required to completely fill the four-fold degenerate surface moir\'e minibands, denoted by the band filling factor, $\nu$, on the right-hand axis (see Methods for definition of $\nu$).

This behavior can be qualitatively captured by a simple model building upon the magnetotransport properties of Bernal graphite. Transport in the moir\'e sample evolves similarly to that of Bernal graphite as the two gate voltages are tuned slightly away from zero, forming a small diagonal resistance stripe. However, as $V_m$ is raised further, the surface moir\'e minibands evolve through a Lifshitz transition corresponding to a sign change in the mass of the free carriers. When $V_m$ corresponds to full filling of the moir\'e band ($\nu=\pm4)$, the surface free charge density returns to approximately zero, and the transport once again mimics the local behavior surrounding $V_{gr} \approx V_m \approx 0$. The sign of the Hall effect similarly flips as the doping at the moir\'e surface switches between electron- and hole-like, corresponding to instances in which Fermi energy crosses a moir\'e band extrema or a Lifshitz transition. For sufficiently large values of $V_{gr}$, the doping of the Bernal graphite surface exceeds the maximum doping possible in the narrow moir\'e band, and the sign of the Hall effect can no longer flip upon further biasing $V_m$.

We verify this picture by calculating the band structure of twisted graphene-graphite with tunable gate-induced surface potentials (see Methods for full details). We adopt the Thomas-Fermi approximation to account for the screening of external electric fields by the graphite bulk~\cite{Koshino2010}. Fig.~\ref{fig:2}c shows the evolution of the band structure for various values of the potential at the moir\'e surface, $U_m$. The Fermi energy is denoted by the black dashed line, and is held fixed at zero energy. We find that gating primarily changes the energy of the moir\'e-like surface states, whereas the graphite-like states in the bulk remain at fixed energy. In contrast, the energy of the moir\'e-like bands remains fixed upon changing the potential at the Bernal graphite surface, $U_{gr}$, as shown in Fig.~\ref{fig:2}d. Instead, only graphite-like states change in energy, resulting in the filling of surface valence or conduction band states depending only on the sign of $U_{gr}$. These calculations support our interpretation of the zig-zag feature observed in transport, since the sign of the free carriers induced at the sample surfaces flips only once upon changing the sign of $U_{gr}$, but inverts repeatedly as the moir\'e surface bands are filled upon changing $U_m$. Extended Data Figures~\ref{fig:band_structure_Um} and~\ref{fig:band_structure_Ug} provide a more detailed analysis of these band structure calculations.

\medskip\noindent\textbf{Hybridized moir\'e and bulk states at high field}

So far, we have found that the charge accumulation layers on the two surfaces do not directly hybridize with each other, and are thus controlled independently by the nearest gate. However, this behavior is known to break down in Bernal graphite at higher magnetic fields~\cite{Yin2019}. In the ultra-quantum limit, only the two lowest (nearly-degenerate) Landau bands cross the Fermi energy, and the electron motion is primarily limited to the $c$-axis~\cite{McClure1968, Yin2019}. Electrons form a set of standing waves that penetrate across the entire bulk owing to the quasi-1D nature of these Landau bands. These states are thus controlled equally by the top and bottom gates. All other Landau bands are gapped within the bulk, and although they can be populated at the surfaces by gating, they form a screening layer and can only be controlled by the nearest gate. As a consequence, we find that Landau fans acquired by sweeping a single gate exhibit two distinct sets of quantum oscillations (QOs). One sequence projects to approximately zero gate voltage at $B=0$, irrespective of the bias applied to the other gate. These QOs correspond to surface-localized states. The second sequence projects to a non-zero gate voltage determined by the bias applied to the opposite gate, in particular following the line of overall charge neutrality. These QOs correspond to states that are extended across the entire bulk (Extended Data Fig.~\ref{fig:graphite_transport}).

The moir\'e surface states in t1+$Z$ graphite form Hofstadter bands at high field~\cite{Hofstadter1976}, which must smoothly evolve into the Bernal graphite Landau bands in the bulk. Despite this additional complexity, our magnetotransport measurements reveal that ungapped bulk states remain extended across the entire sample. We probe this effect by acquiring Landau fan diagrams while sweeping $V_{gr}$ at various fixed values of $V_m$ (Fig.~\ref{fig:3}a). Similar to our observations in Bernal graphite, we see two distinct sequences of QOs that project to different values of $V_{gr}$ at $B=0$. Selected QOs from the sequence corresponding to states localized at the Bernal graphite surface are denoted by black dashed lines, and project to $V_{gr} \approx 0$ irrespective of $V_m$. In contrast, the QOs corresponding to the bulk extended states (denoted by blue dashed lines) project to different values of $V_{gr}$ depending sensitively on $V_m$.

%%%%%%%%%%%%%%%%%%%%%%%%%%%%%%%%%%%%%%%%%%%%%%%%%%%%%%%%%%%%%%%%%%%%%
\begin{figure*}[t]
\includegraphics[width=6.9 in]{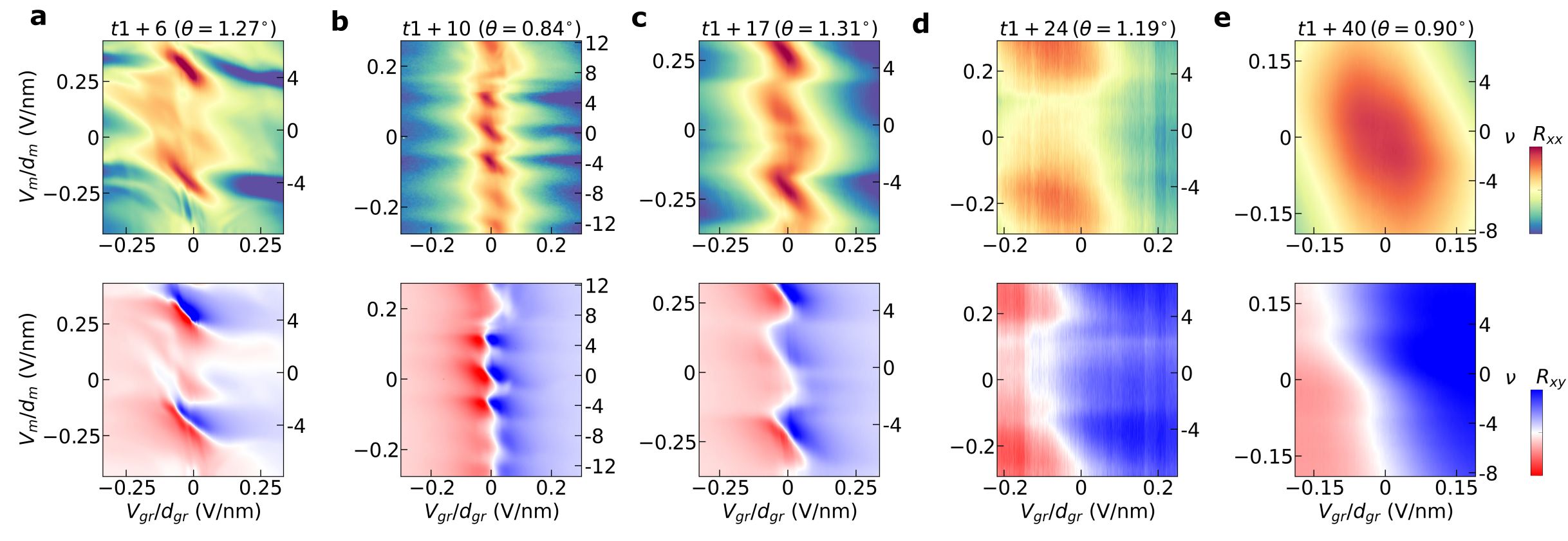} 
\caption{\textbf{Moir\'e modification of graphitic films with varying thickness and twist angle.}
\textbf{a-e}, Maps of the longitudinal (top row) and Hall (bottom row) resistance for different moir\'e graphite devices, acquired at $B=0.5\ \rm{T}$. The graphite thickness and twist angle is indicated above each plot. The extrema of the color scales for each device vary due to the differing graphite thicknesses. The $R_{xx}$ maps are plotted on logarithmic color scales, with extrema of: (\textbf{a}) $50\ \mathrm{\Omega}$ to $1\ \mathrm{k\Omega}$, (\textbf{b}) $5\ \mathrm{\Omega}$ to $500\ \mathrm{\Omega}$ (\textbf{c}) $10\ \mathrm{\Omega}$ to $500\ \mathrm{\Omega}$, (\textbf{d}) $100\ \mathrm{\Omega}$ to $175\ \mathrm{\Omega}$, (\textbf{e}) $50\ \mathrm{\Omega}$ to $150\ \mathrm{\Omega}$. The $R_{xy}$ maps are plotted on linear color scales, with extrema of: (\textbf{a}) $-0.5\ \mathrm{k\Omega}$ to $0.5\ \mathrm{k\Omega}$, (\textbf{b-c}) $-0.3\ \mathrm{k\Omega}$ to $0.3\ \mathrm{k\Omega}$, (\textbf{d-e}) $-50\ \mathrm{\Omega}$ to $50\ \mathrm{\Omega}$. 
}
\label{fig:4}
\end{figure*}
%%%%%%%%%%%%%%%%%%%%%%%%%%%%%%%%%%%%%%%%%%%%%%%%%%%%%%%%%%%%%%%%%%%%%

Figure~\ref{fig:3}c shows the color-coded projections from many Landau fan diagrams overlayed atop the $R_{xy}$ map acquired at $B=0.5$~T. The projection point of the extended states (blue dots) oscillates with $V_m$, tracking closely with the $R_{xy}=0$ contour measured at low field for $-4\leq\nu\leq+4$. These quantities diverge for larger values of $\nu$, however, we still see occasional sharp resets in the projection point of the QOs to values near $V_{gr}=0$ even at larger values of $V_m$ (e.g., at $V_m/d_m\approx-0.2$~V/nm). Overall, this behavior is enabled by the unique nature of the extended standing wave states in Bernal graphite. As illustrated schematically in the top left inset of Fig.~\ref{fig:3}d, the standing wave hybridizes the moir\'e surface states with the graphite bulk states and inextricably mixes the properties of the two. The low- and high-field zig-zag features we see in Fig.~\ref{fig:3}c arise from distinct physical mechanisms, yet exhibit remarkably consistent behavior that is primarily determined by the total gate-induced free charge in the sample.       

Unlike in Bernal graphite, the Landau fan diagrams in twisted grahene-graphite differ substantially depending on which gate is swept. In particular, we see QOs corresponding to the moir\'e bands only upon sweeping $V_{m}$ (Fig.~\ref{fig:3}b). We denote selected QOs projecting to full-filling of the moir\'e surface bands ($\nu=\pm4$) with pink dashed lines, whereas QOs projecting to the charge neutrality point ($\nu=0$) are denoted in purple. Figure~\ref{fig:3}c shows the $B=0$ projection points of these states overlayed atop the $R_{xx}$ map acquired at $B=0.5$~T. Again, we see good agreement between the low- and high-field features. In particular, the low-field transport around $V_{gr}\approx0$ exhibits diagonal resistance features closely matching the evolution of the projection points of the high-field QOs at $\nu=\pm4$ and $0$. We also find that both the resistance and the QO projections depend only weakly on $V_{gr}$ as the bias is raised further. These observations suggest that the moir\'e bands can be doped by changing $V_{gr}$ when the bias is small, but that surface states on the Bernal face screen the effect of changing $V_{gr}$ when the bias is large.

Despite the obvious differences in the Landau fan diagrams acquired by sweeping each gate, Brown-Zak (BZ) oscillations~\cite{Brown1964,Hofstadter1976,Hunt2013,Dean2013,Ponomarenko2013,Kumar2017} appear in both. These occur as maxima in the magnetoconductance, $G_{xx}=R_{xx}/(R_{xx}^2+R_{xy}^2)$, and arise at rational values of the magnetic flux filling of the moir\'e unit cell. The oscillations are most obvious after averaging $G_{xx}$ over the entire range of each gate voltage (denoted as $\langle G_{xx} \rangle$ in Fig.~\ref{fig:3}d), thereby eliminating contributions from individual QOs at a given field. BZ oscillations are anticipated when sweeping $V_m$ (red curve in Fig. \ref{fig:3}d), since charge carriers are directly populating the moir\'e bands. In contrast, the BZ oscillations seen upon sweeping $V_{gr}$ (blue line in Fig. \ref{fig:3}d) are more surprising, since this gate does not directly fill the moir\'e bands. This effect is more clearly visible in a t1+17 device with $\theta=1.31^{\circ}$, which exhibits very sharp magnetoconductance peaks averaged over $V_{gr}$ (top right inset of Fig.~\ref{fig:3}d).

The BZ oscillations correspond to conditions in which charge carriers experience zero effective magnetic field, and thus exhibit straight trajectories in real space~\cite{Kumar2017}. Although it is possible for the moir\'e to propogate through bulk graphite due to structural relaxations, this effect has been found to arise only at ultra-small twist angles~\cite{Halbertal2022}, and is therefore unlikely to be relevant in our samples. Instead, the observation of BZ oscillations upon sweeping $V_{gr}$ indicates that carriers doped into the Bernal graphite surface obey transport properties dictated in part by the moir\'e potential on the opposite surface, providing further evidence of the hybridization of moir\'e and bulk states at high field. We further find that this effect persists in devices where the moir\'e interface is buried at the center of the sample (Extended Data Fig.~\ref{fig:t7_7}). In this case, the moir\'e bands are inaccessible by gating and can not be filled by either gate, but nevertheless generate strong BZ oscillations by hybridizing with the bulk states of each rotated Bernal graphite constituent.

\medskip\noindent\textbf{Discussion}

Our observations appear to be generic for t1+$Z$ graphite, as we see qualitative similarities across different graphite thicknesses and twist angles. Figure~\ref{fig:4} shows $R_{xx}$ and $R_{xy}$ maps acquired at $B=0.5$~T for five samples with Bernal graphite components ranging from $Z=6$ to $40$ layers. The zig-zag resistance feature becomes increasingly obscured for thicker graphite, but we nevertheless see oscillations in $R_{xy}$ that appear to correspond closely with four-fold multiples of $\nu$. The value of $V_m$ required to establish the complete zig-zag feature scales with the twist angle, since bias needed to fully fill the moir\'e surface band is directly proportional to the twist angle. In the high-field regime, we see similar zig-zag evolutions of the Landau fan diagram projections for both the t1+6 and t1+17 graphite samples (Extended Data Figs.~\ref{fig:t1_6} and~\ref{fig:t1_17} and Supplementary Videos 3-5). 

Overall, our results establish a new class of `mixed-dimensional moir\'e materials,' in which a moir\'e potential localized to a single 2D interface fundamentally transforms the properties of an entire bulk crystal. This behavior may generalize to other layered semimetals with low intrinsic bulk doping, in which surface accumulation layers formed by gating can dominate the transport. Similar effects could therefore arise in materials such as WTe$_2$ and ZrTe$_5$. Our work additionally motivates experiments with more complex graphitic structures, including those with moir\'e patterns at both the top and bottom surfaces and those with moir\'e interfaces distributed throughout the bulk of the material. At high magnetic field, standing waves in the bulk may couple these coexisting moir\'e potentials in interesting and exotic new ways. Finally, new complex moir\'e geometries in bulk graphitic thin films may help to unravel the origin of the superconductivity found both in natural few-layer graphene allotropes~\cite{Zhou2021, Zhou2022} and in a growing family of magic-angle twisted graphene structures.  

\section*{Methods}

\textbf{Device fabrication.} Moir\'e devices were fabricated using the ``cut-and-stack'' method~\cite{Chen2019a,Saito2020}. t1+$Z$ structures are made by finding an exfoliated bulk graphite thin film with a connected monolayer graphene region, isolating the two using an atomic force microscope tip, and then stacking one atop the other at the desired twist angle.  The sample with the buried moir\'e was made by isolating two regions from a single 7-layer graphite sheet and stacking them atop each other. All samples were assembled using standard dry-transfer techniques with a polycarbonate (PC)/polydimethyl siloxane (PDMS) stamp~\cite{Wang2013}. All devices are encapsulated in flakes of BN and graphite, and then transferred onto a Si/SiO$_2$ wafer. The temperature was kept below 180$^{\circ}$C during device fabrication to preserve the intended twist angle. The number of graphite layers in each device, $Z$, was determined by atomic force microscopy measurements after encapsulation. Standard electron beam lithography, CHF$_3$/O$_2$ plasma etching, and metal deposition techniques (Cr/Au) were used to define the complete stack into a Hall bar geometry~\cite{Wang2013}.

\textbf{Transport measurements.} Transport measurements were performed in a Cryomagnetics variable temperature insert, and were conducted in a four-terminal geometry with a.c. current excitation of 10-200 nA using standard lock-in techniques at a frequency of 17.7 Hz. Some of the measurements acquired for the Supplementary Videos 1-2 for the t1+10 device were performed in a Bluefors dilution refrigerator at a nominal base temperature of 20 mK.

\textbf{Twist angle determination.} The twist angle $\theta$ is determined by fitting the sequence of QOs arising upon sweeping $V_m$. The charge carrier density required to fill the moir\'e superlattice is given by $n_s=8\theta^2/\sqrt{3}a^2$, where $a=0.246$~nm. The value of $n_s$ is determined by tracing the QOs corresponding to full filling of the moir\'e surface bands to $B=0$. The band filling factor, $\nu$, is defined such that $\nu=\pm4$ at doping $\pm n_s$, where the numerical factor of 4 corresponds to the spin and valley degeneracy of graphene.

QOs projecting to full filling ($\nu=\pm4$) are evident in all devices shown in Fig.~\ref{fig:4} except for the t1+40 device. The twist angle for the t1+40 device was instead estimated through piezoelectric force microscopy (PFM) imaging of the moir\'e pattern (Extended Data Fig.~\ref{fig:pfm}). PFM was performed on the transfer slide during the sample fabrication directly after picking up the 40-layer graphite and the monolayer graphene~\cite{McGilly2020}. The twist angle is extracted by calculating the average of the three moir\'e lattice points in the Fourier transform of the PFM image. The twist angle in the t1+6 device was also independently confirmed in this manner. 

When observed, the BZ oscillation sequence provides an independent measure of the twist angle. Magnetoconductance peaks are expected to occur when the magnetic flux $\phi=4B/n_s$ is equal to $p/q$ times the flux quantum $\phi_0=h/e$, where $h$ is Planck's constant and $p, q$ are integers. We extract $n_s$ by fitting the observed peaks to a series of rational $\phi/\phi_0=1/2, 1/3, 1/4,$ etc. In all devices, we find that this value agrees with that extracted by tracking the QOs to within a few percent.

\textbf{Temperature dependence measurements.} We track the evolution of the low-field magnetotransport in the t1+17 device in Extended Data Fig.~\ref{fig:temp_dep}. We see signatures of the zig-zag feature persisting up to at least 50 K. As the temperature is raised further, the structure in the $R_{xx}$ map becomes washed out. $R_{xy}$ continues to exhibit a sign change, but along a straight line with a slope that becomes more vertical in the map with higher temperature. At room temperature, transport is nearly unaffected by changing $V_m$, potentially due to lower mobility on the moir\'e surface compared to the Bernal graphite surface.

\textbf{Transport model of Bernal graphite at low field.} We capture the low-field transport behavior of Bernal graphite with a four-component Drude model. Magnetotransport is characterized by the conductivity tensor $\bm{\sigma}$, with a corresponding current density $\bm{j}=\bm{\sigma}\bm{E}$ under an electric field $\bm{E}$. Each of the four carrier species are independent, and have a two-dimensional carrier density denoted as $n_i$, where $i$ is $t$, $b$, $e$, or $h$. In order, these correspond to the charges on the top and bottom graphite surfaces, and the intrinsic electron and hole free carriers in the graphite bulk. Each carrier species has an associated mobility, $\mu_i$. In the absence of a magnetic field, the conductivity is a scalar and the contribution from the $i$-th carrier is $\sigma_{i}=en_i \mu_i$. In the presence of a magnetic field, the conductivity tensor is $\bm{\sigma}=\sum_i \bm{\sigma}_i$ where the contribution from the $i$-th carrier is
\begin{align*}
    \bm{\sigma}_i = \frac{\sigma_{i}}{1+(\mu_i B)^2} \left(
    \begin{array}{cc}
       1  & \eta_i \mu_i B \\
     -\eta_i \mu_i B   & 1
    \end{array}
    \right),
\end{align*}
with $\eta_i=\mp 1$ for electrons and holes, respectively. The resistivity tensor is $\bm{\rho}=\sigma^{-1}$, where the diagonal and off-diagonal elements separately represent the longitudinal ($\rho_{xx}$) and transverse ($\rho_{xy}$) resistivities.

Graphite is a nearly compensated semimetal, and for simplicity we assume $n_e=n_h$. We take previously measured parameters of bulk graphite~\cite{Soule1958,McClure1958}, in which the three-dimensional bulk carrier density is $n_{\rm{3D}}=3\times10^{18}\ \rm{cm^{-3}}$ and the mobility is $\mu_e=\mu_h=1\times10^6 \ \rm{cm^2/Vs}$. We assume the same value for the surface mobilities, $\mu_t$ and $\mu_b$. The bulk electron and hole carrier densities correspond to a two-dimensional density of $n_{\rm{2D}}=2\times10^{12}\ \rm{cm^{-2}}$ per graphene sheet. To model our 23-layer Bernal graphite sample, we assign 21 layers of $n_{\rm{2D}}$ as the fixed bulk density, $n_e=n_h=21\times n_{\rm{2D}}=4\times10^{13}\ \rm{cm^{-2}}$. We then vary the surface density of the two remaining layers, corresponding to changing the top and bottom gate voltages over typical experimentally accessible values. 

Extended Data Fig. \ref{fig:graphite_theory}a shows the results of the calculation, which qualitatively match the experimental observations in Fig. \ref{fig:2}a. In particular, we see the largest resistivity along the line of overall charge neutrality, with a corresponding change in the sign of the Hall effect. We also see that the resistivity is largest when both of the surfaces are undoped, also consistent with our experimental results. The qualitative agreement between experiment and theory does not depend strongly on the precise values of the graphite parameters we assume. However, as an additional check, we repeat the calculation with unrealistically large bulk density, $n_e=n_h=2\times10^{15}\ \rm{cm^{-12}}$ (Extended Data Fig. \ref{fig:graphite_theory}b). In this case, we find that virtually no gate dependence can be observed in the longitudinal resistivity, which changes by only tenths of an ohm, compared to over a hundred ohms in the calculation performed with realistic graphite parameters. These calculations therefore establish that the gate-tunable transport we observe in graphite arises as a consequence of its modest intrinsic bulk doping.

\textbf{Band structure calculation.} We first calculate the Hamiltonian for $Z$-layer Bernal graphite.
We define $|\bm{k}, A_l \rangle$ and $|\bm{k}, B_l \rangle$ as the Bloch states at the $K$ point of layer $l$.
By arranging the basis as $|\bm{k}, A_1 \rangle,|\bm{k}, B_1 \rangle;|\bm{k}, A_2 \rangle,|\bm{k}, B_2 \rangle;...;|\bm{k}, A_Z \rangle,|\bm{k}, B_Z \rangle;$, the Hamiltonian around the $K$ point is given by
\begin{equation*}
H_{\rm ZG}(\bm{k})=
\left(
\begin{array}{cccc}
H_{\rm D}(\bm{k}) - U_1     &        \Gamma(\bm{k})   &   0                        & \\
 \Gamma^{\dagger}(\bm{k})   & H_{\rm D}(\bm{k}) - U_2 &   \Gamma^{\dagger}(\bm{k}) & \\
 0                          &        \Gamma(\bm{k})   &    H_{\rm D}(\bm{k})- U_3  & \\
                            &                         &                            & \ddots\\
\end{array}
\right),
\end{equation*}
with
\begin{equation*}
H_{\rm D}(\bm{k})=
\left(
\begin{array}{cc}
               0                             & \frac{\sqrt{3}}{2}\gamma_0 (k_x - i k_y) \\
\frac{\sqrt{3}}{2}\gamma_0 (k_x + i k_y) &                    0              \\
\end{array}
\right),
\end{equation*} and
\begin{equation*}
\Gamma(\bm{k})=
\left(
\begin{array}{cc}
-\frac{\sqrt{3}}{2}\gamma_4 (k_x - i k_y)      & -\frac{\sqrt{3}}{2}\gamma_3 (k_x + i k_y) \\
                     \gamma_1                  & -\frac{\sqrt{3}}{2}\gamma_4 (k_x - i k_y) \\
\end{array}
\right),
\end{equation*}
where $U_j$ is the electronic potential at $j$th layer. 
We set the parameters $(\gamma_0,\gamma_1,\gamma_3,\gamma_4) = (2.6,0.36,0.28,0.14)$~eV~\cite{Charlier1991}.

To calculate the band structure of twisted graphene-graphite, we further place a single graphene layer on the $Z$-layer graphite and twist it by $\theta$.
The effective Hamiltonian can be written as
\begin{equation*}\label{eq_Ham_all}
H=
\left(
\begin{array}{cc}
H_{\rm MG} & H_{\rm int}^{\dagger} \\
H_{\rm int} & H_{\rm ZG} \\
\end{array}
\right),
\end{equation*}
where $H_{MG}$ is the Hamiltonian for the twisted monolayer graphene.
$H_{int}$ is the interlayer coupling term between the monolayer and the topmost layer of the graphite.
We define $|\bm{k}', A_0 \rangle$ and $|\bm{k}', B_0 \rangle$ as the Bloch states of the twisted graphene at $R(\theta)K$, where $R(\theta)$ is the rotation matrix in the $xy$-plane at an angle $\theta$.
By using this basis, the Hamiltonian for the graphene can be represented as
\begin{equation*}
H_{\rm MG}=
\left(
\begin{array}{cc}
                     U_0                   &  \frac{\sqrt{3}}{2}\gamma_0(k_x' - i k_y') \\
\frac{\sqrt{3}}{2}\gamma_0 (k_x' + i k_y') &                             U_0              \\
\end{array}
\right).
\end{equation*}
The Bloch wavevector of the twisted graphene and the $Z$-layer graphite are coupled when $\bm{k} = \bm{k}' + \bm{q}_j$ $(j=0,1,2)$, where $\bm{q_0}=(0,0)$, $\bm{q}_1=\frac{1}{L_M}(-\frac{2\pi}{\sqrt{3}},-2\pi)$ and $\bm{q}_2=\frac{1}{L_M}(\frac{2\pi}{\sqrt{3}},-2\pi)$, where $L_M$ is the moir\'e lattice constant.
Under these constraints, the interlayer coupling is given by
\begin{equation*}
H_{int}= \sum_{j=0}^2 t_{int}
\left(
\begin{array}{cc}
    \alpha          &e^{-i\frac{2\pi j}{3}} \\
e^{i\frac{2\pi j}{3}} &         \alpha      \\
\end{array}
\right),
\end{equation*}
and is otherwise zero.
We take the parameters $t_{int} = 0.11$~eV and $\alpha=0.5$~\cite{Koshino2020}, the latter of which captures the effects of lattice relaxations.

We additionally consider the effective model for $Z$-layer graphite twisted atop $Z$-layer graphite, such that the moir\'e interface is buried at the center of the material.
The Hamiltonian is constructed following same analysis as above, but replacing $H_{\rm MG}$ with $H_{\rm ZG}(\bm{k}')$.
The two rotated graphite constituents are coupled by $H_{\rm int}$ at their rotated interface. 

We capture the effects of gating by adopting the Thomas-Fermi approximation to account for the screening of external electric fields by the graphite bulk~\cite{Manes2007,Koshino2010}.
For small gate voltages, the electric field is given by
\begin{equation*}
\label{eq_Ham_mono}
U_j = U_m e^{-jd/\lambda_s} + U_{gr} e^{-(Z-j)d/\lambda_s} \quad(j=0,...,Z)
\end{equation*}
where $\lambda_s\approx1.3d$, and $d$ is the interlayer distance.
$U_{m/gr}$ is the potential for the \moire/graphite surface.

Finally, we numerically calculate the eigenvalues and eigenstates by truncating at a sufficiently large momentum and diagonalizing the Hamiltonian.
The electron of the $n$-th eigenvalue is distributed along the graphite $c$-axis as follows:
\begin{equation*}
\label{eq_Ham_mono}
\langle z_{n\bm{k}} \rangle = \sum_{l}\sum_{\sigma=A,B} z_l |\phi_{n\bm{k},\sigma_l}|^2
\end{equation*}
where the eigenstate is written as $| \psi_{n\bm{k}} \rangle = \sum_{ \sigma, l} \phi_{n\bm{k},\sigma_l} | \bm{k} ,\sigma_l \rangle$ and $z_l = 2 (z_0 - l d) / L_z$.
$L_z$ and $z_0$ denote the width and the center of the system, respectively.

\section*{Data Availability}
Source data are available for this paper. All other data that support the plots within this paper and other findings of this study are available from the corresponding author upon reasonable request.

\section*{acknowledgments}
We thank D. Cobden, V. Fal'ko, S. Slizovskiy, and M. Rudner for valuable discussions. This work was supported by NSF CAREER award DMR-2041972 and NSF MRSEC 1719797. M.Y. acknowledges support from the State of Washington funded Clean Energy Institute. D.W. was supported by an appointment to the Intelligence Community Postdoctoral Research Fellowship Program at University of Washington administered by Oak Ridge Institute for Science and Education (ORISE) through an interagency agreement between the U.S. Department of Energy and the Office of the Director of National Intelligence (ODNI). E.T. and E.A-M. were supported by NSF GRFP DGE-2140004. Y.R. and D.X. were supported by the Department of Energy, Basic Energy Sciences, Materials Sciences and Engineering Division, Pro-QM EFRC (DE-SC0019443). M.F. was supported by JST CREST Grant Number JPMJCR20T3, Japan and a JSPS Fellowship for Young Scientists. K.W. and T.T. acknowledge support from the Elemental Strategy Initiative conducted by the MEXT, Japan (Grant Number JPMXP0112101001) and JSPS KAKENHI (Grant Numbers 19H05790, 20H00354 and 21H05233). This research acknowledges usage of the millikelvin optoelectronic quantum material laboratory supported by the M. J. Murdock Charitable Trust.\\

\section*{Author contributions}
D.W., E.T. and E.A-M. fabricated the devices and performed the measurements. M.F. performed the band structure calculations. Y.R. performed the magnetotransport calculation. T.C. and D.X. supervised the calculations. K.W. and T.T. grew the BN crystals. D.W., E.T., E.A.-M. and M.Y. analyzed the data and wrote the paper with input from all authors.

\section*{Competing interests}
The authors declare no competing interests.

\section*{Additional Information}
Correspondence and requests for materials should be addressed to M.Y.

\section*{Supplementary Information}
Supplementary Videos 1-5.

\bibliographystyle{naturemag}
\bibliography{references}

\clearpage

\renewcommand{\figurename}{Extended Data Fig.}
\renewcommand{\thesubsection}{S\arabic{subsection}}
\setcounter{secnumdepth}{2}
%\renewcommand{\theequation}{S\arabic{equation}}
%\renewcommand{\thetable}{S\arabic{table}}
%\subsubsectionfont{\normalfont\large\itshape\underline}
\setcounter{figure}{0} 
\setcounter{equation}{0}

\onecolumngrid
\newpage
\section*{Extended Data}

\begin{figure*}[h]
\includegraphics[width=6.9 in]{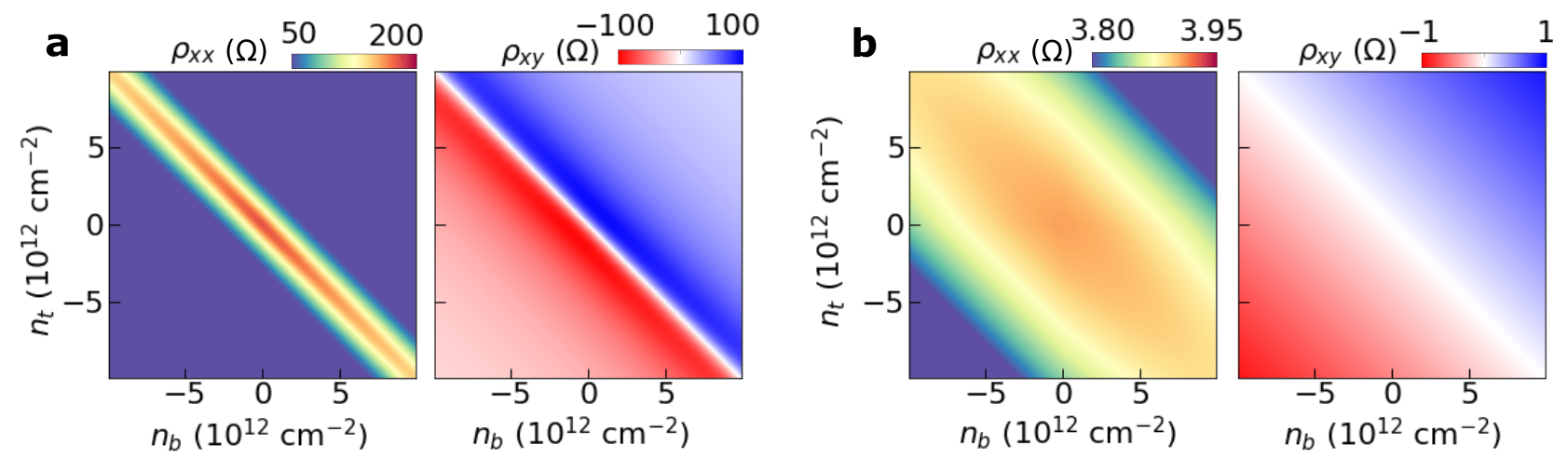} 
\caption{\textbf{Four-component Drude transport model of Bernal graphite at $B=0.5$~T.}
\textbf{a} Calculation of $\rho_{xx}$ (left) and $\rho_{xy}$ (right) for 23-layer graphite with $n_e=n_h=21\times n_{\rm{2D}}=4\times10^{13}\ \rm{cm^{-2}}$ and $\mu=1\times10^6 \ \rm{cm^2/Vs}$. Full details of the calculation are described in the Methods section.
\textbf{b}, The same calculation with unrealistically large values of the intrinsic graphite doping, $n_e=n_h=1\times10^{13}\ \rm{cm^{-2}}$. The change in resistivity with gating is orders of magnitude smaller than in \textbf{a}, indicating that the gate-tunability arises owing to the semimetallic nature of graphite.}
\label{fig:graphite_theory}
\end{figure*}

\begin{figure*}[h]
\includegraphics[width=6.9 in]{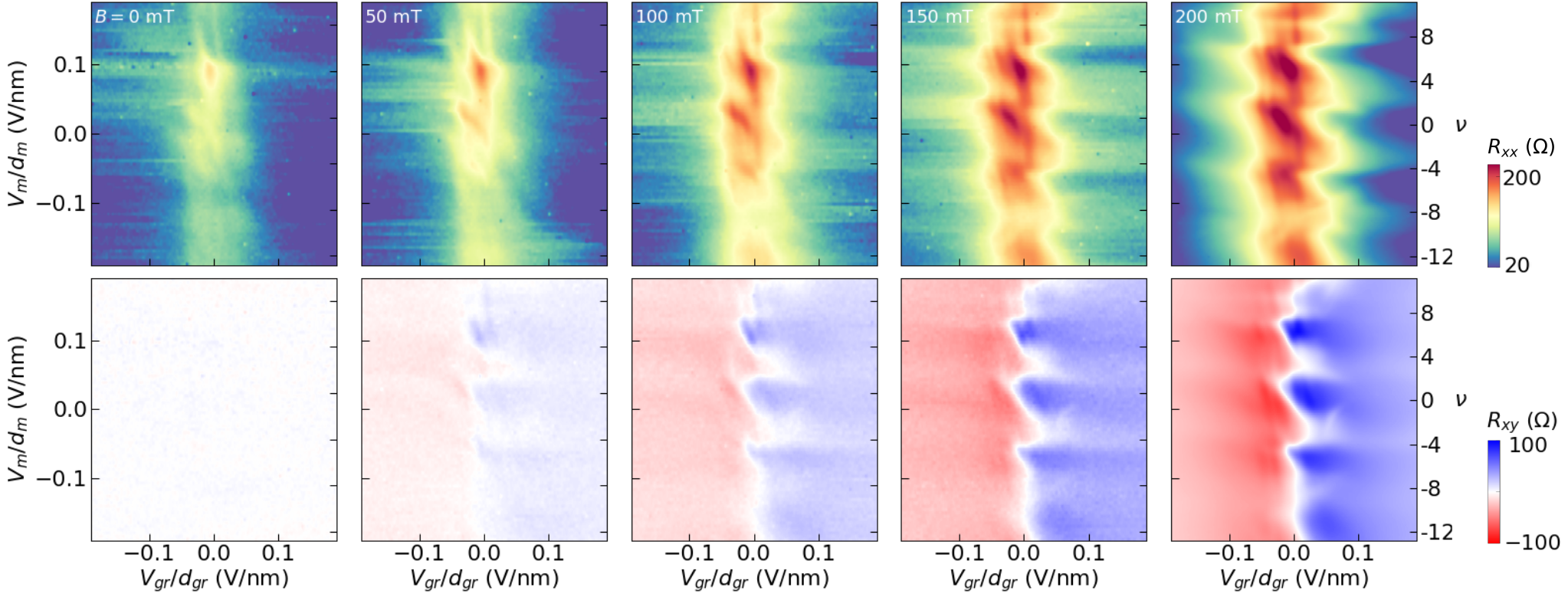} 
\caption{\textbf{Low-field evolution of transport in the t1+10 device.}
Longitudinal (top) and Hall (bottom) resistance measurements acquired in steps of $B=50$~mT, as indicated in the top left of each column. The zig-zag transport behavior first becomes evident at fields as low as 50 mT, and becomes more obvious as the field is raised.}
\label{fig:t1_10_lowB}
\end{figure*}

\begin{figure*}[h]
\includegraphics[width=6.9 in]{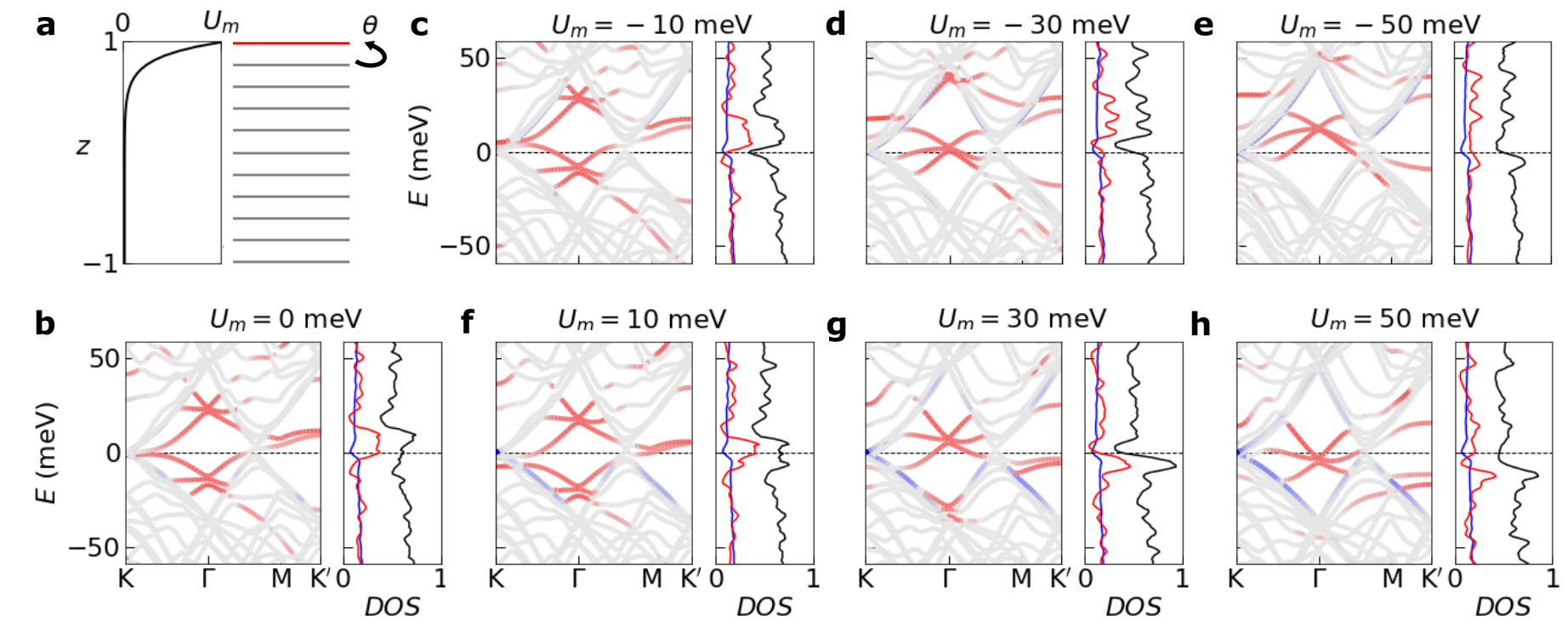} 
\caption{\textbf{Band structure calculations with varying $U_m$ for the $t1+10$ device.}
\textbf{a}, The assumed potential, $U_m$, as a function of $\rm{z}$, given by a Thomas-Fermi screening approximation. The constituent graphene layers are indicated schematically to the right of the plot. The red layer denotes the monolayer graphene sheet twisted to $\theta=0.84^{\circ}$ with respect to the Bernal graphite layers beneath (grey lines). 
\textbf{b}-\textbf{h}, Calculated band structure (left) and density of states integrated over the moir\'e Brillouin zone (right) for selected values of $U_m$, as  denoted above each panel. $U_{gr}=0$ for all panels. The bands are color-coded following the same scheme as in Fig.~\ref{fig:2}c of the main text. The black curve shows the total density of states. The blue shows the density of states filtered by $\langle \rm{z} \rangle<-0.6$, corresponding to the Bernal graphite surface states. The red line shows the same but for $\langle \rm{z} \rangle>0.6$, corresponding to moir\'e surface states. The normalization of the density of states is the same in all panels.}
\label{fig:band_structure_Um}
\end{figure*}

\begin{figure*}[h]
\includegraphics[width=6.9 in]{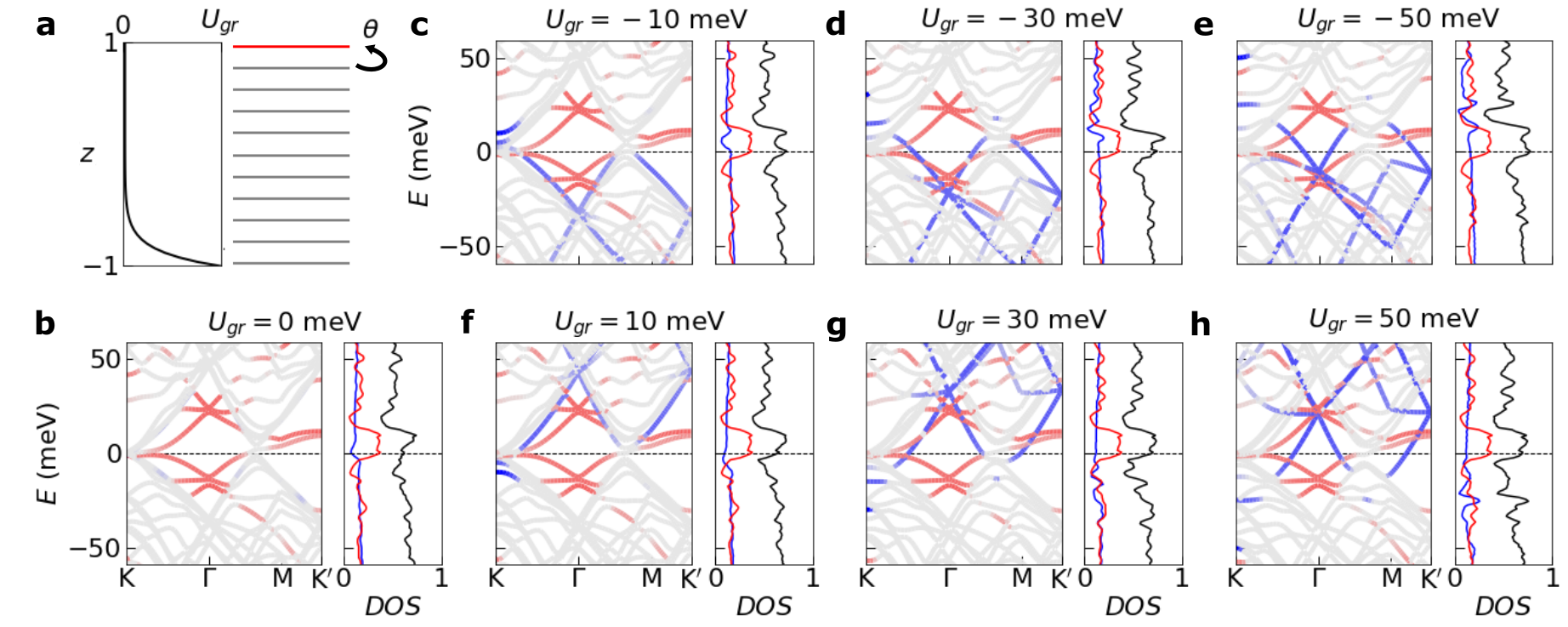} 
\caption{\textbf{Band structure calculations with varying $U_{gr}$ for the $t1+10$ device.}
Same as Extended Data Fig. \ref{fig:band_structure_Um}, but instead varying the potential on the graphite surface, $U_{gr}$, with $U_m=0$.} 
\label{fig:band_structure_Ug}
\end{figure*}

\begin{figure*}[h]
\includegraphics[width=6.9 in]{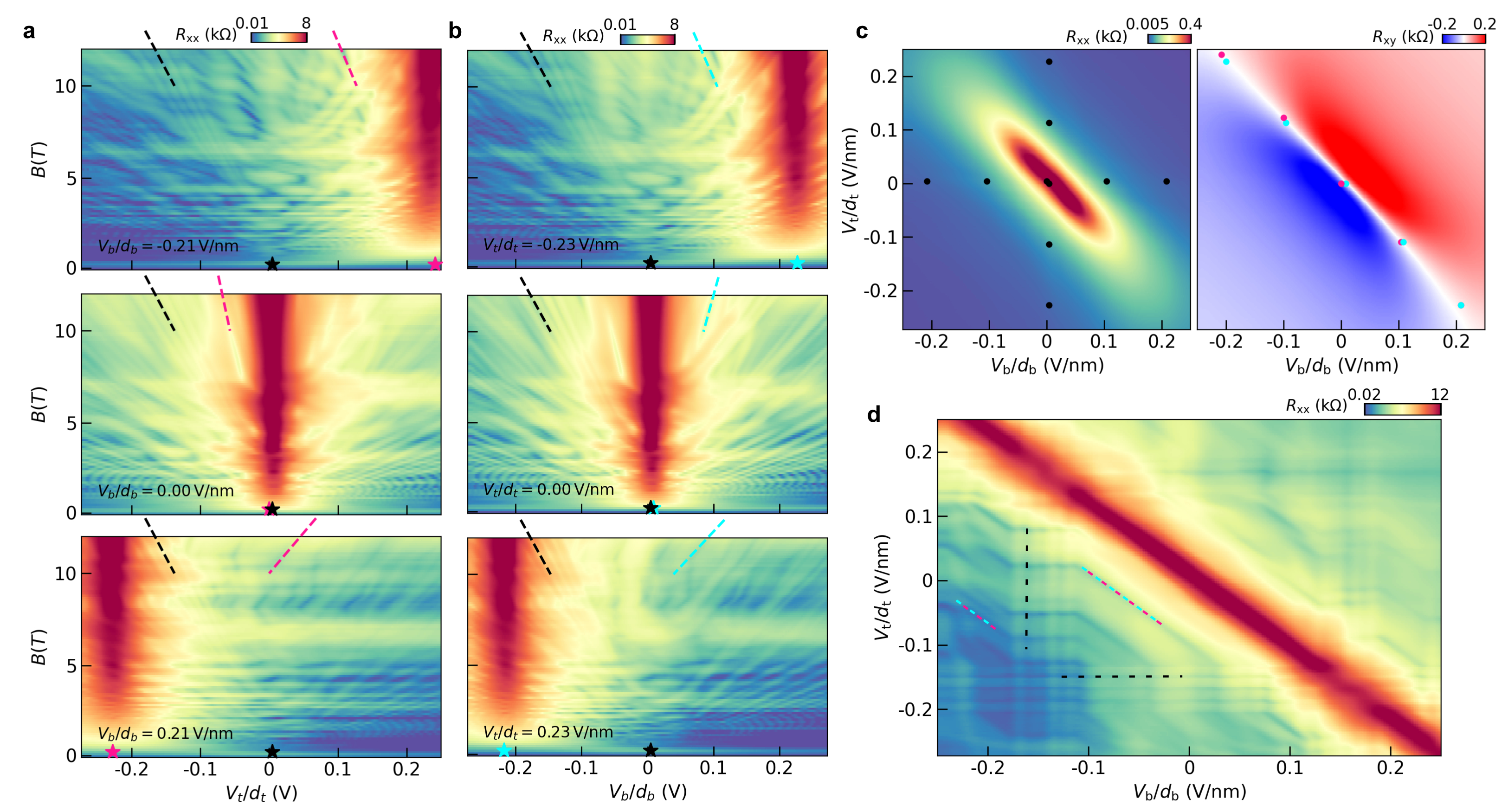} 
\caption{\textbf{High-field transport in Bernal graphite.}
\textbf{a}, Landau fan diagrams from a 24-layer graphite device (different from the one shown in the main text) acquired by sweeping the top gate voltage, $V_{t}$, at various fixed values of the bottom gate voltage, $V_{b}$, as indicated in each panel. The black dashed lines denote selected QOs that project to $V_{t} \approx 0$ at $B=0$, and the pink lines denote QOs that project to a $V_{t}\neq0$ that depends on the value of $V_b$.
\textbf{b}, Similar Landau fans, but with fixed $V_t$ and sweeping $V_b$. The QOs projecting to $V_{b}\neq0$ are denoted in blue.
\textbf{c}, Longitudinal (left) and Hall (right) resistance maps acquired at $B=0.5$~T. The QOs projecting to approximately zero gate voltage in each Landau fan, corresponding to surface-localized states, are overlayed on the $R_{xx}$ map and form a cross. The QOs projecting to non-zero gate voltages, corresponding to extended bulk states, are overlayed on the $R_{xy}$ map and closely track the condition of overall charge neutrality.
\textbf{d}, $R_{xx}$ map acquired at $B=12$~T. Dashed black lines denote selected QOs that depend only on a single gate, which arise from localized states on either the top or bottom graphite surfaces. These correspond to the QOs denoted in black in the Landau fans. The blue/pink dashed lines denote QOs that depend on both gates, which evolve parallel to the line of overall charge neutrality. These states correspond to the QOs denoted in blue and pink in the Landau fans, and arise from the extended bulk states.}
\label{fig:graphite_transport}
\end{figure*}

\begin{figure*}[h]
\includegraphics[width=\textwidth]{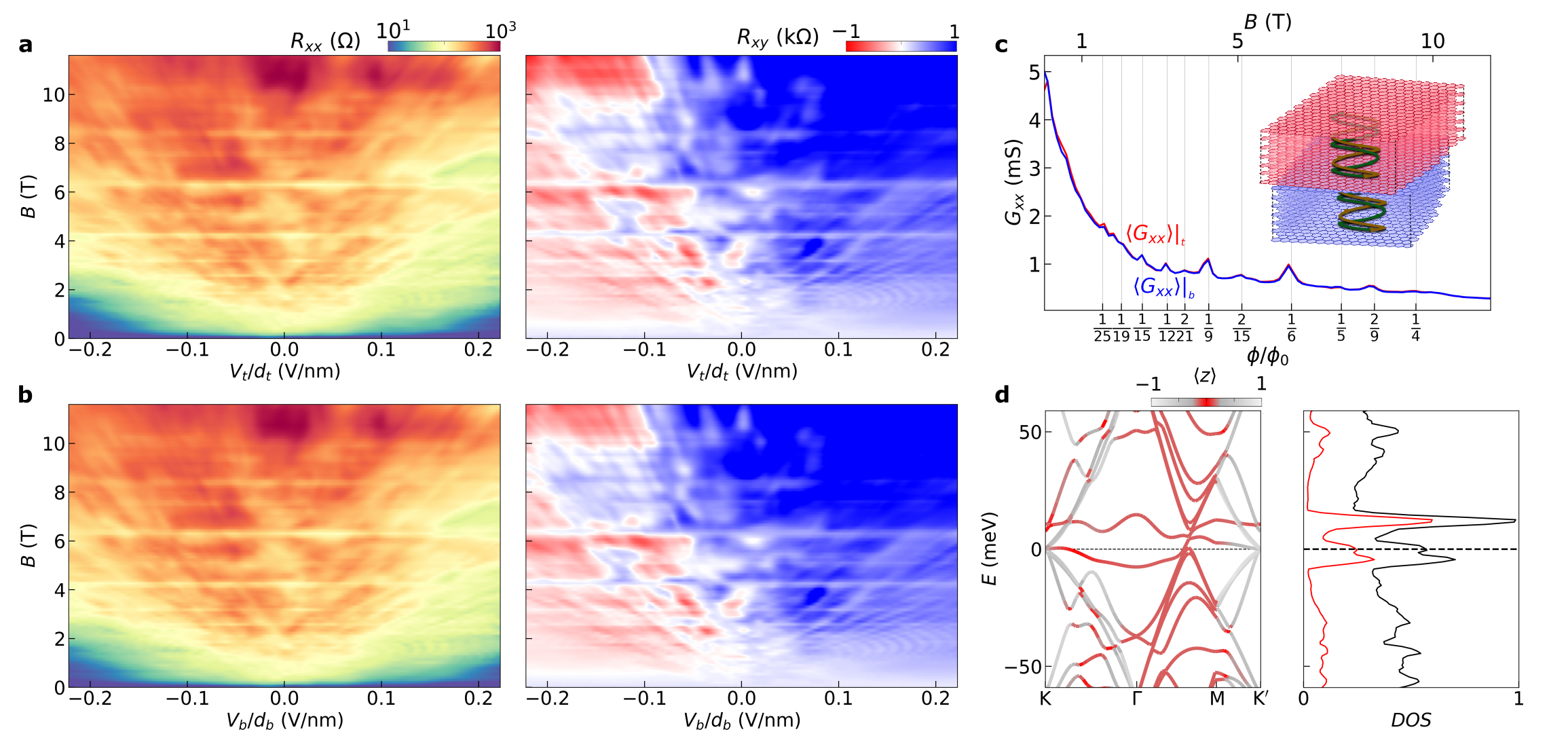} 
\caption{\textbf{High-field transport in a device with a buried \moire interface.}
\textbf{a}, $R_{xx}$ (left) and $R_{xy}$ (right) Landau fans acquired as $V_t$ is swept with $V_b=0$ in a device consisting of 7 layers of Bernal graphite stacked and rotated atop another 7 graphite layers with $\theta=1.26^{\circ}$.
\textbf{b}, Same as (\textbf{a}), but sweeping $V_b$ with $V_t=0$. The fans acquired by sweeping each gate are nearly identical to one another owing to the symmetry of the structure. Since the moir\'e is buried, we do not observe signatures of moir\'e band filling in either Landau fan. However, both fans show clear horizontal features corresponding to Brown-Zak oscillations arising due to the buried moir\'e. 
\textbf{c}, Magnetoconductance averaged across all gate voltages for the top (red curve) and bottom (blue curve) gates. These are nearly identical, and both display a clear sequence of Brown-Zak oscillations. The inset shows a cartoon schematic in which separate standing waves couple the top and bottom bulk graphite states to the buried moir\'e interface. The hybridization of the buried moir\'e with the bulk states is required to generate the BZ oscillations seen in transport.
\textbf{d}, (Left) Band structure calculation of this structure showing a moir\'e band localized at the center of the twisted graphitic thin film. The color scale is defined as in Fig.~\ref{fig:2}c of the main text. In this case, the moir\'e bands are found at $\langle \rm{z} \rangle \approx 0$, since the moir\'e is located at the center of the structure. (Right) The density of states integrated over the moir\'e Brillouin zone. The red filtered curve corresponds to the four central graphene sheets, whereas the black corresponds to the total density of states.}
\label{fig:t7_7}
\end{figure*}

\begin{figure*}[h]
\includegraphics[width=7 in]{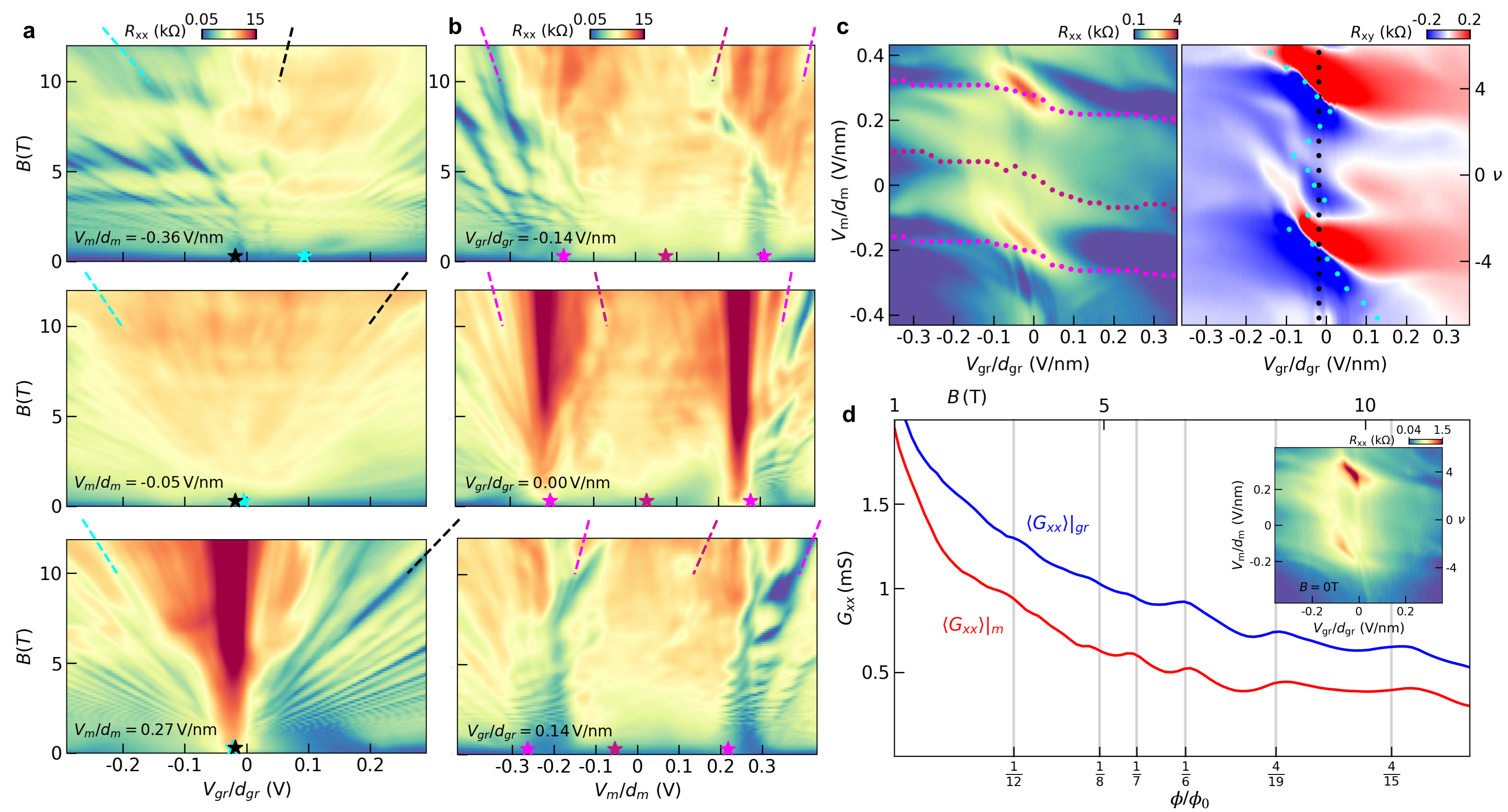} 
\caption{\textbf{High-field transport in a t1+6 device with $\theta=1.27^{\circ}$.}
\textbf{a}, Landau fan diagrams acquired by sweeping $V_{gr}$ at various fixed values of $V_{m}$. The black dashed lines denote selected QOs that project to $V_{gr} \approx 0$ at $B=0$. The blue lines denote selected QOs that project to a $V_{gr}\neq0$ that depends on $V_{m}$. 
\textbf{b}, Landau fan diagrams acquired by sweeping $V_{m}$ at fixed values of $V_{gr}$. The purple (pink) dashed lines denote selected QOs that project to $\nu=0$ ($\nu=\pm 4$). 
\textbf{c}, Longitudinal (left) and Hall (right) resistance maps acquired at $B=0.5$~T. Zero-field projections of the $\nu=0$ and $\pm4$ states from the $V_m$ Landau fans are overlayed on the $R_{xx}$ map. Zero-field projections of QOs from the $V_{gr}$ Landau fans are overlayed on the $R_{xy}$ map. 
\textbf{d}, Conductance, $G_{xx}$, as a function of magnetic field. The blue curve is averaged over all values of $V_{gr}$ for the Landau fan in \textbf{a} acquired at $V_{m}/d_m= -0.05\,\rm{V/nm}$. The red curve is averaged over a range of $V_m$ values corresponding to $|\nu|<4$ for the Landau fan in \textbf{b} acquired at $V_{gr}= 0$. Brown-Zak oscillations case be seen upon sweeping either gate. (Inset) Longitudinal resistance map acquired at $B=0\ \rm{T}$.}
\label{fig:t1_6}
\end{figure*}

\begin{figure*}[h]
\includegraphics[width=7 in]{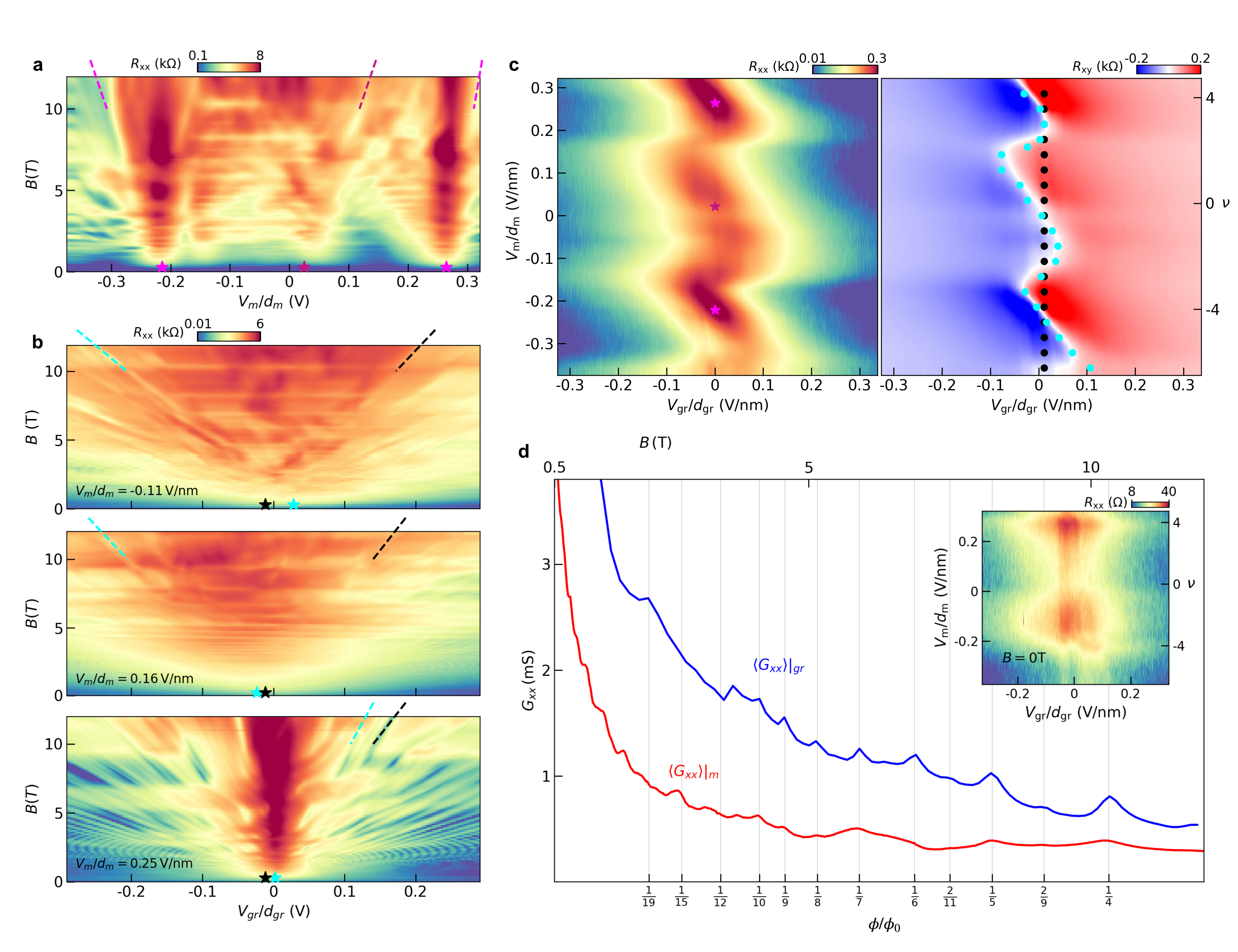}
\caption{\textbf{High-field transport behavior in a t1+17 device with $\theta=1.31^{\circ}$.}
\textbf{a}, Landau fan diagram acquired by sweeping $V_{m}$ at $V_{gr}=0$. The purple (pink) dashed lines denote selected QOs that project to $\nu=0$ ($\nu=\pm 4$). 
\textbf{b}, Landau fan diagrams acquired by sweeping $V_{gr}$ at various fixed values of $V_{m}$. The black dashed lines denote selected QOs that project to $V_{gr} \approx 0$ at $B=0$. The blue lines denote selected QOs that project to a $V_{gr}\neq0$ that depends on $V_{m}$. 
\textbf{c}, Longitudinal (left) and Hall (right) resistance maps acquired at $B=0.5$~T. Zero-field projections of the $\nu=0$ and $\pm4$ states from the $V_m$ Landau fans are overlayed on the $R_{xx}$ map. Zero-field projections of QOs from the $V_{gr}$ Landau fans are overlayed on the $R_{xy}$ map. 
\textbf{d}, Conductance, $G_{xx}$, as a function of magnetic field. The blue curve is averaged over all values of $V_{gr}$ for the Landau fan in (\textbf{a}) acquired at $V_{m}/d_m= -0.05\,\rm{V/nm}$. The red curve is averaged over a range of $V_m$ values corresponding to $|\nu|<4$ for the Landau fan in (\textbf{b}) acquired at $V_{gr}= 0$. Brown-Zak oscillations case be seen upon sweeping either gate. (Inset) Longitudinal resistance map acquired at $B=0\ \rm{T}$.
}
\label{fig:t1_17}
\end{figure*}

\begin{figure*}[h]
\includegraphics[width=5 in]{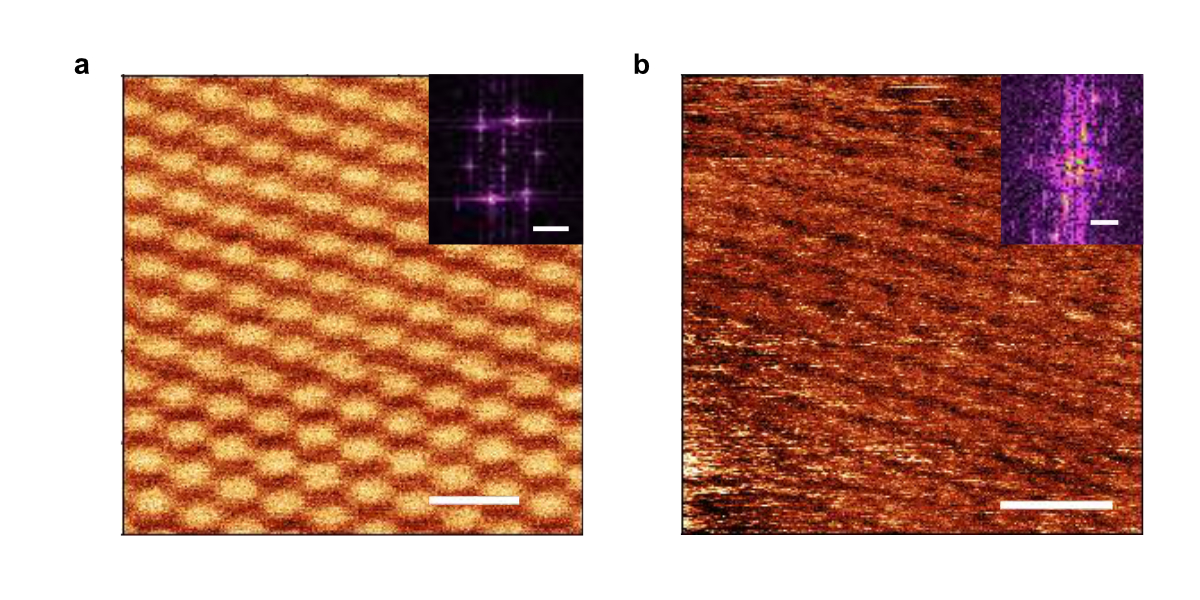} 
\caption{\textbf{Twist angle determination by PFM.}
\textbf{a}-\textbf{b} Piezoelectric force microscopy (PFM) imaging of (a) t1+6 and (b) t1+40 structure performed directly after stacking the monolayer graphene atop the bulk graphite (i.e., while the sample is still on the transfer slide, before BN encapsulation). The twist angle is extracted by calculating the average of the three moir\'e lattice points in the Fourier transform of the PFM image, shown in the insets. The estimated twist angles for the t1+6 and t1+40 devices are $1.43\pm0.15^{\circ}$ and $0.90\pm0.16^{\circ}$ respectively. The t1+40 device appears to have a large amount of heterostrain, causing a distortion of the usual hexagonal shape of the moir\'e lattice points. Scale bars correspond to: (\textbf{a}) 20 nm (inset, 0.1 nm$^{-1}$), (\textbf{b}) 50 nm (inset, 0.1 nm$^{-1}$).}
\label{fig:pfm}
\end{figure*}

\begin{figure*}[h]
\includegraphics[width=7 in]{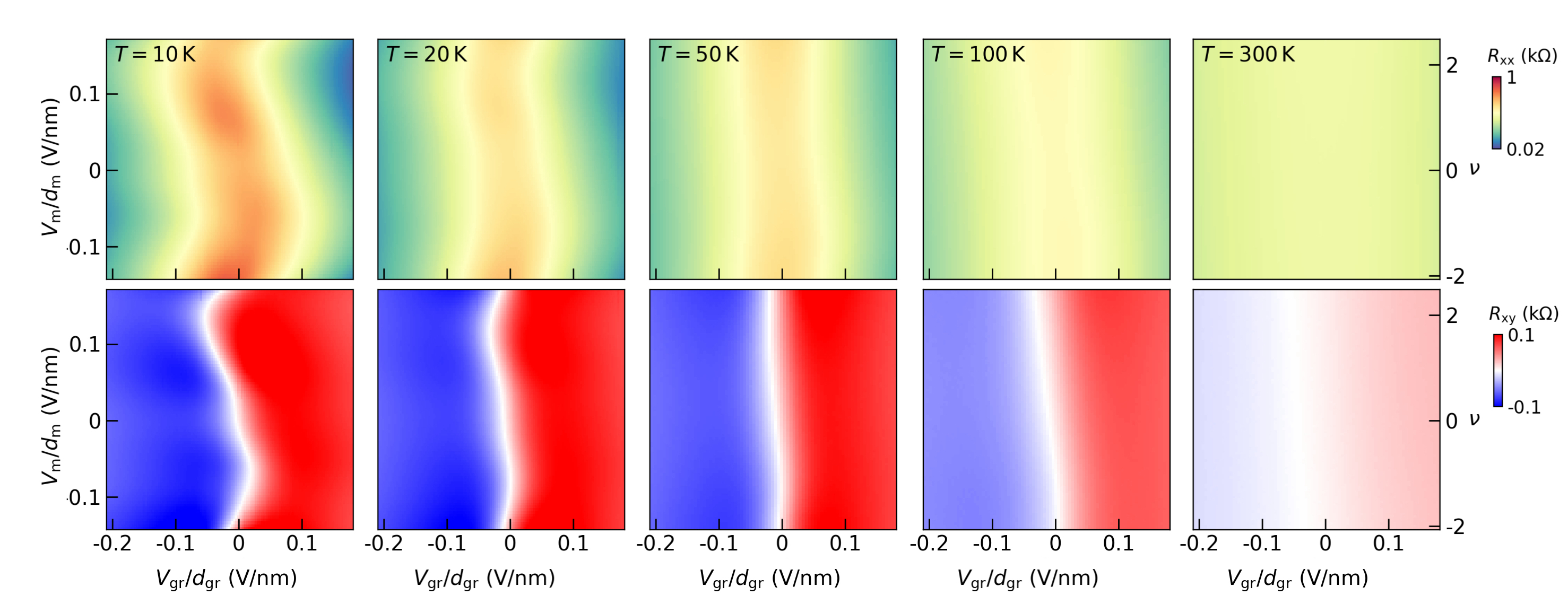} 
\caption{\textbf{Temperature dependence of the transport in the t1+17 device.}
Maps of the longitudinal (top) and Hall (bottom) resistance at $B=0.5$~T, acquired at various temperatures as indicated at the top left of each panel. The zig-zag behavior persists weakly up to approximately 50 K, above which $R_{xx}$ becomes nearly featureless and $R_{xy}$ becomes nearly independent of $V_m$.}
\label{fig:temp_dep}
\end{figure*}

\clearpage
\newpage
\section*{Supplementary Videos}

\textbf{Supplementary Video 1.} (left) Landau fan diagrams from the t1+10 device acquired by sweeping $V_{gr}$ at the indicated values of $V_{m}$. Selected QOs are denoted at the top of the image by black and blue arrows, with the corresponding projection points at $B=0$ indicated by black and blue dots. Some arrows correspond to QOs that are only visible at intermediate values of $B$ in the map (i.e., QOs that do not directly connect to the arrow itself). We note that there are many other visible QOs not indicated by arrows, all of which appear to project either to the black or blue dots. The black and blue dots are chosen by judging the best fit for the entire sequence of visible QOs. In general, the blue dots align with the value of $V_{gr}$ corresponding to the highest resistance over a wide range of $B$ in the map. (right) $R_{xy}$ map acquired at $B=0.5$~T with the corresponding black and blue dots overlayed.\\

\textbf{Supplementary Video 2.} (left) Landau fan diagrams from the t1+10 device acquired by sweeping $V_m$ at the indicated values of $V_{gr}$. Selected QOs are denoted at the top of the image by pink and purple arrows, with the corresponding projection points at $B=0$ indicated by pink and purple dots. Some arrows correspond to QOs that are only visible at intermediate values of $B$ in the map (i.e., QOs that do not directly connect to the arrow itself). We note that there are many other visible QOs not indicated by arrows, all of which appear to project the pink and purple dots. The pink and purple dots are chosen by judging the best fit for the entire sequence of visible QOs. (right) $R_{xx}$ map acquired at $B=0.5$~T with the corresponding pink ($\nu=0$) and purple ($\nu=\pm4)$ dots overlayed.\\

\textbf{Supplementary Video 3.} Same as Supplementary Video 1, but for the t1+6 device.\\

\textbf{Supplementary Video 4.} Same as Supplementary Video 2, but for the t1+6 device.\\

\textbf{Supplementary Video 5.} Same as Supplementary Video 1, but for the t1+17 device. This video only includes Landau fans with $V_{gr}/d_{gr}\geq -0.18\ \rm{V/nm}$. The fans acquired for $V_{gr}/d_{gr} < -0.18\ \rm{V/nm}$ were taken only up to $B=5\ \rm{T}$ due to technical constraints in those particular measurements, and are not included in the video for the sake of continuity. Nevertheless, these lower-field fans still enable unambiguous QO projections, as plotted on the $R_{xx}$ map.

\end{document}